\documentclass[a4paper]{article}

\usepackage{ISCSLP2022}

\usepackage{graphicx}
\usepackage{amsmath}
\usepackage{amssymb}
\usepackage{booktabs}
\usepackage{multirow}
\usepackage{footmisc}
\usepackage[hyphens]{url}

\usepackage{hyperref}
\hypersetup{
colorlinks=true,
linkcolor=blue,
citecolor=blue,
urlcolor=blue,
}

\title{Speech-enhanced and Noise-aware Networks for Robust Speech Recognition}
\name{Hung-Shin Lee$^1$, Pin-Yuan Chen$^1$, Yao-Fei Cheng$^1$, Yu Tsao$^2$, and Hsin-Min Wang$^1$}
\address{
$^1$Institute of Information Science, Academia Sinica\\
$^2$Research Center for Information Technology Innovation, Academia Sinica}
\email{hungshinlee@gmail.com, freddy@iis.sinica.edu.tw}

\begin{document}

\maketitle
\begin{abstract}
Compensation for channel mismatch and noise interference is essential for robust automatic speech recognition. Enhanced speech has been introduced into the multi-condition training of acoustic models to improve their generalization ability. In this paper, a noise-aware training framework based on two cascaded neural structures is proposed to jointly optimize speech enhancement and speech recognition. The feature enhancement module is composed of a multi-task autoencoder, where noisy speech is decomposed into clean speech and noise. By concatenating its enhanced, noise-aware, and noisy features for each frame, the acoustic-modeling module maps each feature-augmented frame into a triphone state by optimizing the lattice-free maximum mutual information and cross entropy between the predicted and actual state sequences. On top of the factorized time delay neural network (TDNN-F) and its convolutional variant (CNN-TDNNF), both with SpecAug, the two proposed systems achieve word error rate (WER) of 3.90\% and 3.55\%, respectively, on the Aurora-4 task. Compared with the best existing systems that use bigram and trigram language models for decoding, the proposed CNN-TDNNF-based system achieves a relative WER reduction of 15.20\% and 33.53\%, respectively. In addition, the proposed CNN-TDNNF-based system also outperforms the baseline CNN-TDNNF system on the AMI task.
\end{abstract}
\noindent\textbf{Index Terms}: robust speech recognition, autoencoder, multi-condition training, noise-aware training

\section{Introduction}
\label{sec:intro}

For nearly a decade, the development of deep learning-based methods for automatic speech recognition (ASR) has advanced significantly, regardless of whether the acoustic model (AM) is trained using a deep neural network/hidden Markov model (DNN/HMM)-based process or a phone-free end-to-end (E2E) structure. In realistic situations, the speech input to an ASR system may include a considerable amount of interference from various types of background noise, and there may be a channel mismatch between it and the training speech. Therefore, it is very important to improve the robustness of ASR. This study is focused on boosting the noise robustness of Gaussian mixture model (GMM)/DNN-based ASR.

As described in \cite{Fan2021}, there are three deep learning-based approaches that can boost the robustness of ASR. The first approach uses speech enhancement (SE) models with various neural structures, such as the denoising autoencoder (DAE) \cite{Vincent2008,Fujimoto2019}, dense neural network \cite{Narayanan2013}, and generative adversarial network (GAN) \cite{Pascual2017,Fan2019}, to pre-process the input speech. These SE models are usually trained to estimate the ideal ratio mask (IRM) for speech or directly restore clean speech from noisy speech. However, the training objective is not directly related to the ultimate goal of ASR, which is to minimize phone or word errors. Therefore, the robustness of the downstream ASR system largely depends on the performance of SE components \cite{Han2015,Wang2016}.

The second approach is multi-condition training (MCT), which uses clean and noisy speech for acoustic modeling. The purpose of MCT is to optimize the familiarity of the ASR system with various types of noisy speech to avoid mismatches between training and testing environments. Although MCT still suffers from unexpected conditions and speech distortion \cite{Seltzer2013,Li2014}, in several typical ASR competitions, such as the CHiME challenge, it has been widely used through data augmentation with noise addition or reverberation simulation \cite{Manohar2019}. In addition to noisy speech, the enhanced speech generated by SE models can be added to the training set \cite{Fujimoto2019}. As a special case of MCT, SpecAug directly applies certain spectrogram deformations to speech signals for data augmentation, which significantly improves the ASR performance \cite{Park2019}.

The third approach is joint training (JT), in which an SE model consisting of an autoencoder \cite{Gao2015,Mimura2016} or GAN \cite{Liu2018} is attached to the AM to minimize the losses of SE and ASR in a unified manner. The ASR-related loss is mainly associated with connectionist temporal classification or attention-based sequence-to-sequence models in E2E systems \cite{Liu2019,Soni2019,Fan2021}. As for the GMM/DNN topology, only the phoneme/state-level cross entropy (CE) has been adopted in the literature so far \cite{Gao2015,Narayanan2015,Wang2016,Mimura2016}. JT compensates for the shortcomings of the first approach in several ways. By keeping the contributions of recognition and enhancement in balance, it prevents the enhanced speech from being over-smoothed. Enhanced speech itself is a good source of feature-level data augmentation. For example, Wang and Wang concatenated noisy mel-frequency cepstral coefficient (MFCC) features and enhanced filter-bank features as the input of the back-end AM \cite{Wang2016}, whereas Fan \textit{et al.} proposed a gated recurrent fusion method that fuses noisy and enhanced features to address speech distortion \cite{Fan2021}. Moreover, in \cite{Gao2015,Mimura2016,Liu2019,Soni2019}, only the enhanced features were used as the input of AM. In this case, the SE loss functions as a sample-level regularizer to avoid overfitting.

In addition to the above-mentioned methods with the help of SE, noise-aware training (NAT), which appends ``noise vectors'' to the input feature sequence, has also attracted attention in robust ASR \cite{Seltzer2013,Kundu2016,Raj2020}. The main difference between them is whether the noise vectors are estimated through noise detection \cite{Seltzer2013,Raj2020} or directly derived from noisy speech \cite{Kundu2016}. Noise representations in
NAT is learnable using neural networks. NAT can also be thought of as a type of feature-level data augmentation.

In this paper, we propose a unified framework that combines MCT, JT, and NAT to achieve robust ASR. The framework is implemented on GMM/DNN AMs trained using the lattice-free maximum mutual information (LF-MMI) criterion \cite{Povey2016}. As shown in Table \ref{tab:comparison}, the main novelty of this study is twofold. First, the simultaneous optimization of the frame-level cross-entropy, utterance-level LF-MMI, and acoustic feature reconstruction error is unprecedented in JT-based robust ASR. Second, the introduction of noise-aware training and data augmentation with noise not only strengthens the capability of the enhancement component to extract purer enhanced speech, but also enables the subsequent AM to utilize the estimated noise feature to filter out the noise in noisy speech more effectively.

\begin{table}
\centering
\caption{JT-based methods for robust ASR. EnhC and Enh represent the enhancement component and the enhanced feature, respectively. G/D denotes the GMM/DNN topology.}
\small
\vspace{-5pt}
\setlength{\tabcolsep}{4pt}
\label{tab:comparison}
\begin{tabular}{cccc}
\toprule
\textbf{Method} & \textbf{Topo.} & \textbf{EnhC} & \textbf{Augment.} \\
\midrule
\multicolumn{1}{l}{Gao \textit{et al.} (2015) \cite{Gao2015}} & G/D & DAE & - \\
\multicolumn{1}{l}{Mimura \textit{et al.} (2016) \cite{Mimura2016}} & G/D & DAE & - \\
\multicolumn{1}{l}{Wang \textit{et al.} (2016) \cite{Wang2016}} & G/D & T/F IRM & Enh \\
\multicolumn{1}{l}{Kundu \textit{et al.} (2016) \cite{Kundu2016}} & G/D & FL & Enh \& Noise \\
\multicolumn{1}{l}{Liu \textit{et al.} (2018) \cite{Liu2018}} & E2E & GAN & - \\
\multicolumn{1}{l}{Liu \textit{et al.} (2019) \cite{Liu2019}} & E2E & T/F PSA & - \\
\multicolumn{1}{l}{Soni \textit{et al.} (2019) \cite{Soni2019}} & G/D & T/F IRM & - \\
\multicolumn{1}{l}{Fan \textit{et al.} (2021) \cite{Fan2021}} & E2E & T/F IAM & Enh \\
\midrule
\textbf{Ours} & G/D & DAE & Enh \& Noise \\
\bottomrule
\end{tabular}
\vspace{-10pt}
\end{table}

\section{Proposed Models}

Fig. \ref{fig:structure} illustrates the schematic diagram of our proposed neural architecture, where two additional components and AM are sequentially connected and trained jointly. This work has been open-sourced at \url{https://github.com/Sinica-SLAM/kaldi-SENAN}.

\subsection{Acoustic Model}
The AM converts a frame representation $\mathbf{x}_{in}$ into triphone-state scores. The input $\mathbf{x}_{in}$ is the concatenation of three types of features: noisy features and the i-vector $\mathbf{x}_{nsy}$, enhanced features $\mathbf{x}_{enh}$, and noise-aware features $\mathbf{x}_{nse}$.
\begin{equation}
\mathbf{x}_{in}=\mathbf{x}_{nsy}\oplus\mathbf{x}_{enh}\oplus\mathbf{x}_{nse},
\label{eq:concat}
\end{equation}
where $\oplus$ denotes the concatenation operator.

\begin{figure}[t]
\centering
\includegraphics[width=0.48\textwidth]{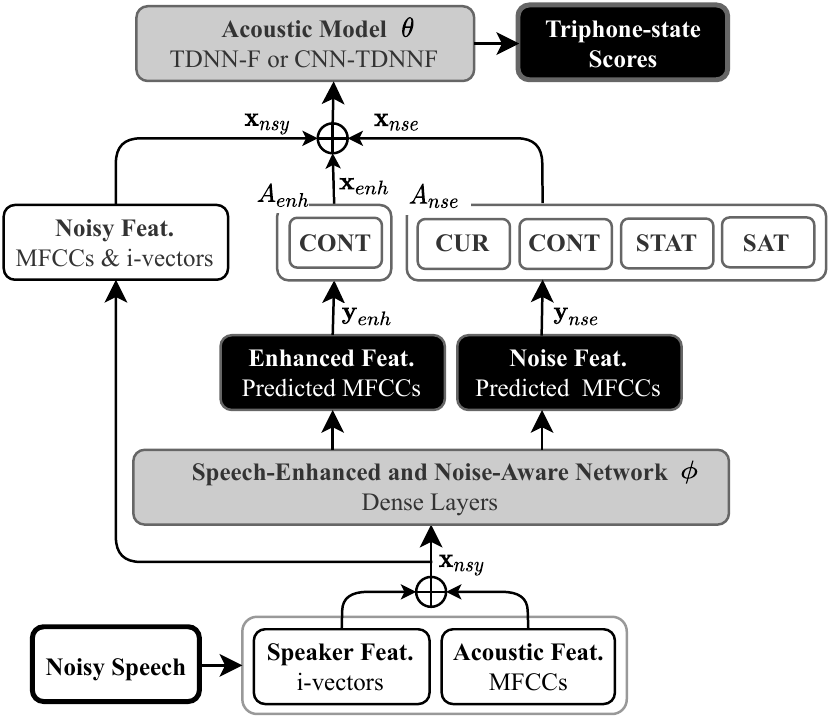}
\vspace{-15pt}
\caption{The proposed speech-enhanced and noise-aware acoustic model, where the three black blocks are directly related to loss functions during training.}
\label{fig:structure}
\vspace{-15pt}
\end{figure}

The AM, parameterized by $\theta$, generates the triphone-state scores (see the top black block in Fig. \ref{fig:structure}). In this study, the AM was implemented on a factorized time-delay neural network (TDNN-F) or its variant with convolutional layers (CNN-TDNNF) \cite{Peddinti2015a,Peddinti2015b,Povey2018}. We derived two ASR-related objective functions. The first is the CE between the predicted triphone-state scores and ground truth defined by
\begin{equation}
\mathcal{L}_{CE} = -\sum_{u=1}^{U} \sum_{t=1}^{T_u} \log p_{\theta}(\mathbf{s}^{ut}|\mathbf{x}^{ut}_{in}),
\label{eq:ce}
\end{equation}
where $U$ is the number of training utterances, $T_u$ is the number of frames of utterance $u$, and $\mathbf{s}^{ut}$ and $x^{ut}_{in}$ are the true triphone-state label and acoustic input of the $t$-th frame of utterance $u$, respectively. The label is derived by GMM-based forced alignment. The second is based on MMI, a discriminative criterion designed to maximize the probability of the reference transcription while minimizing the probabilities of all other transcriptions \cite{Vesely2013}. As in \cite{Hadian2018b}, the LF-MMI is defined as
\begin{equation}
\mathcal{F}_{LF-MMI} = \sum_{u=1}^{U} \log \dfrac{p_{\theta}(\mathbf{x}^{u}_{in}|M_{\mathbf{w}_u})P(M_{\mathbf{w}_u})}{p_{\theta}(\mathbf{x}^{u}_{in}|M_{den})},
\label{eq:lf_mmi}
\end{equation}
where $\mathbf{w}_u$ is the reference word sequence of utterance $u$. The composite HMM graph $M_{\mathbf{w}_u}$ represents all possible state sequences pertaining to $\mathbf{w}_u$ and is called the numerator graph. $M_{den}$ is the denominator graph that models all the possible sequences. It has traditionally been estimated using lattices because the full denominator graph can become large and significantly slow the computation. Povey \textit{et al.} derived MMI training of DNN/HMM models using a full denominator graph (hence, the name lattice-free) \cite{Povey2016}.

\subsection{Speech-enhanced and Noise-aware Network (SENAN)}
\label{sec:senan}

If $\mathbf{x}_{enh}$ and $\mathbf{x}_{nse}$ in Eq. \ref{eq:concat} are discarded, our proposed model is reduced to the conventional MCT-based framework. The speech-enhanced and noise-aware network (SENAN) $\phi$, constructed with dense layers, decomposes the frame-wise noisy features $\mathbf{x}_{nsy}$ into the speech-enhanced and noise-aware parts, i.e., $\mathbf{y}_{enh}$ and $\mathbf{y}_{nse}$. Then, the aggregation functions output $\mathbf{x}_{enh}$ and $\mathbf{x}_{nse}$ from $\mathbf{y}_{enh}$ and $\mathbf{y}_{nse}$ by
\begin{equation}
[\mathbf{x}_{enh}, \mathbf{x}_{nse}]=[A_{enh}(\mathbf{y}_{enh}), A_{nse}(\mathbf{y}_{nse})],
\label{eq:y_to_x}
\end{equation}
where $[\mathbf{y}_{enh}, \mathbf{y}_{nse}]=\phi(\mathbf{x}_{nsy})$, and $A(\cdot)$ denotes the aggregation function, which is used to leverage the contextual or statistical information near the working frame. As shown in Fig. \ref{fig:structure}, the aggregation function can consider the current frame only (CUR), the current frame appended with one preceding and one following frame (CONT), the concatenation of two statistical measurements (the mean and variance vectors) derived from a context window of 150 frames (STAT) \cite{Medennikov2021}, and single-head self-attention in a local range of five preceding and two following frames around the current frame (SAT) \cite{Vaswani2017}. Based on preliminary results, we use CONT as the aggregation function for $\mathbf{y}_{enh}$, and evaluate the four functions for $\mathbf{y}_{nse}$ in this paper.

To ensure that the enhanced features $\mathbf{y}_{enh}$ and estimated noise $\mathbf{y}_{nse}$ are as similar as possible to the corresponding clean speech and true noise, respectively, we minimize the mean squared error (MSE) defined by
\begin{equation}
\mathcal{L}_{MSE}^{\{enh,nse\}} = \sum_{u=1}^{U} \sum_{t=1}^{T_u} \left|\left| \mathbf{y}_{\{enh,nse\}}^{ut}-\mathbf{\hat{y}}_{\{clean,nse\}}^{ut}\right|\right|^{2}_{2},
\label{eq:mse}
\end{equation}
where $\left|\left|\cdot \right|\right|^{2}_{2}$ is the 2-norm operator. $\mathbf{y}^{ut}_{enh}$ is the predicted enhanced MFCC vector, and $\mathbf{\hat{y}}^{ut}_{clean}$ is the corresponding MFCC vector extracted from clean speech. $\mathbf{y}^{ut}_{nse}$ is the predicted noise MFCC vector, and $\mathbf{\hat{y}}^{ut}_{nse}$ is the corresponding MFCC vector extracted from the true noise signal.

\subsection{Final Objective Function}

The final objective function $L$ to be minimized by backpropagation is a combination of ASR-related, speech-enhanced, and noise-aware losses, and is expressed as
\begin{equation}
\mathcal{L} = \alpha \mathcal{L}_{CE}-\mathcal{F}_{LF-MMI}+\beta (\mathcal{L}^{enh}_{MSE}+\mathcal{L}^{nse}_{MSE}),
\label{eq:final}
\end{equation}
where $\alpha$ and $\beta$ are weighting factors. In previous studies in \cite{Yang2017,Huang2019,Lee2020}, only the CE and MSE metrics were jointly optimized in AM training. In our experiments, we set $\alpha$ to 5 as in most previous studies and heuristically set $\beta$ to 0.2.

\section{Experimental Settings}

We used the Kaldi toolkit \cite{Povey2011} to implement our models, and evaluated them on two datasets: the Aurora-4 corpus \cite{Parihar2002} and the AMI meeting corpus \cite{Carletta2006}.

\subsection{Datasets and Derivation of Noise}
Aurora-4 \cite{Parihar2002} is a medium-level vocabulary task. The transcriptions were based on the Wall Street Journal corpus (WSJ0) \cite{Paul1992}. The dataset contains 16 kHz speech data with additive noise and linear convolutional channel distortion, which were synthetically introduced into clean speech. The training set contains 7,138 utterance pairs from 83 speakers, where each pair consists of clean speech and speech corrupted by one of six different noises at 10--20 dB SNR. The test set was generated by the same types of noise and microphones and was grouped into four subsets: clean, noisy, clean with channel distortion, and noisy with channel distortion, which are referred to as A, B, C, and D, respectively. The speech data were recorded by two microphones, one of which was a near-field microphone. The speech recorded by this microphone can be considered clean and noise-free. In contrast, the speech recorded by the second microphone can be considered noisy speech. According to the assumption that noisy speech can be represented by the sum of a speech signal and a noise signal \cite{Boll1979}, we adjusted the volume of the noisy speech and subtracted the clean speech from the noisy speech to obtain the corresponding noise as the ground truth.

The AMI task is more challenging than the Aurora-4 task. The AMI corpus \cite{Carletta2006} contains approximately 77 h of meeting recordings, including 107 k utterances for training and 12.6 k utterances (8.7 h) for testing. The speech signals were captured and synchronized by multiple microphones, including individual head microphones, lapel microphones, and one or more microphone arrays. In this task, we took the speech recorded by individual head microphones and the standard SDM (single distant mic.) as the clean speech and the corresponding noisy speech, respectively. The method used for obtaining the noise was the same as that used for the Aurora-4 task.

\subsection{Data Preparation and Preprocessing}

Our training data (noisy and clean speech) were first organized for standard GMM training/alignment to derive the frame-wise triphone-state ground truth. To avoid overfitting the neural network to improve its robustness, speed and volume perturbations \cite{Ko2015} were used to triple the amount of the training data. The trigram language model (LM) provided by the Kaldi recipe with a closed 5 k vocabulary was used in the Aurora-4 task, whereas the trigram LM trained with the transcriptions of the training set was used in the AMI task.

To extract acoustic features, spectral analysis was applied to 25-ms speech frames every 10 ms. For each frame, 40 high-resolution MFCCs were derived by discrete cosine transform conducted on 40 mel-frequency bins. Utterance-level mean subtraction was used for feature normalization. The noisy features were formed by concatenating the acoustic features and the 100-dimensional speaker-related i-vector \cite{Dehak2011}.

\subsection{Neural Structures}
\subsubsection{Speech-enhanced and noise-aware network}

We used the multi-task autoencoder (MTAE) in \cite{Zhang2018} to construct the SENAN $\phi$ in Fig. \ref{fig:structure}. MTAE has triangular-shaped shared units, and is a unified structure with a denoising autoencoder and a ``de-speeching'' autoencoder. There are 1,024 and 2,048 nodes in the first and last hidden layers, respectively. The number of nodes increases linearly from low to high hidden layers, and the number of hidden layers is five.

\subsubsection{Acoustic model}

One of the best neural structures for realizing the AM $\theta$ in Fig. \ref{fig:structure} is the factorized form of TDNN \cite{Peddinti2015b}, called TDNN-F. TDNN-F and TDNN have the same structure, but the layers of the former are compressed by singular value decomposition and trained by random initialization with semi-orthogonal constraints in one of the two factors of each matrix \cite{Povey2018}. Our TDNN-F was constructed with 13 hidden layers, each containing 1,024 hidden nodes and 128 linear bottleneck nodes. The parameters of the final layer were factorized using a 192-node linear bottleneck layer. The mini-batch sizes were set to 128 and 64. The initial and final effective learning rates were set to 0.01 and 0.001, respectively, and the total number of training epochs was set to 20. In our CNN-TDNNF model, there were six convolutional layers followed by nine TDNN-F layers. The time offsets and height offsets were set to ``-1, 0, 1'' in terms of Kaldi for all convolutional layers, and the number of filters in the six convolutional layers was set to 48, 48, 64, 64, 64, and 128, respectively. The parameters of the TDNN-F layers and hyperparameters were set to the same value as in the TDNN-F system. The total number of training epochs was set to 10.

In addition, we applied SpecAug to the input of the AM during training. It involves deformation in two dimensions: temporal and feature-level masking.

\begin{table}[t!]
\centering
\caption{WERs (\%) on Aurora-4 with respect to different aggregation functions for the predicted noise features. AM: TDNN-F without SpecAug.}
\vspace{-5pt}
\label{tab:agg}
\begin{tabular}{cccccc}
\toprule
\multicolumn{1}{c}{\textbf{Aggregation Func.}} &
\multicolumn{1}{c}{\textbf{A}} & 
\multicolumn{1}{c}{\textbf{B}} & 
\multicolumn{1}{c}{\textbf{C}} & 
\multicolumn{1}{c}{\textbf{D}} & 
\multicolumn{1}{c}{\textbf{Avg.}} \\
\midrule
\midrule
CURR & 1.61 & 3.59 & 3.12 & 9.88 & 6.11 \\
CONT & 1.87 & 3.53 & 3.34 & 9.61 & 6.00 \\
STAT & 1.74 & 3.29 & 3.01 & 9.60 & \textbf{5.86} \\
ATTN & 1.77 & 3.41 & 3.03 & 9.53 & 5.89 \\
\bottomrule
\end{tabular}
\vspace{-15pt}
\end{table}

\section{Experimental Results}

\subsection{Primary Results}

\begin{table}[t!]
\centering
\caption{WERs (\%) on Aurora-4 with respect to TDNN-F and CNN-TDNNF. -/+ denote without/with, respectively. Proposed$^\dagger$ is the oracle case, where SENAN is removed and the enhanced/noise features (i.e., $\mathbf{y}_{enh}$/$\mathbf{y}_{nse}$ in Eq. \ref{eq:y_to_x}) are replaced by $\mathbf{\hat{y}}_{clean}$/$\mathbf{\hat{y}}_{nse}$ derived from clean speech and true noise.}
\vspace{-5pt}
\label{tab:aurora4}
\begin{tabular}{@{\hspace{0pt}}c@{\hspace{6pt}}c@{\hspace{6pt}}ccccc}
\toprule
\multicolumn{2}{c}{\textbf{TDNN-F}} &
\multicolumn{1}{c}{\textbf{A}} & 
\multicolumn{1}{c}{\textbf{B}} & 
\multicolumn{1}{c}{\textbf{C}} & 
\multicolumn{1}{c}{\textbf{D}} & 
\multicolumn{1}{c}{\textbf{Avg.}} \\
\midrule
\multirow{3}{*}{-SpecAug} & Baseline & 1.74 & 3.83 & 3.29 & 10.31 & 6.42 \\
\cmidrule{2-7}
& Proposed & 1.74 & 3.29 & 3.01 & 9.60 & \textbf{5.86} \\
\cmidrule{2-7}
& Proposed$^\dagger$ & 1.55 & 1.60 & 1.63 & 1.62 & 1.61 \\
\cmidrule{1-7}
\multirow{2}{*}{+SpecAug} & Baseline & 1.33 & 2.55 & 2.15 & 6.28 & 4.03 \\
\cmidrule{2-7}
& Proposed & 1.59 & 2.45 & 2.30 & 6.01 & \textbf{3.90} \\
\midrule
\midrule
\multicolumn{2}{c}{\textbf{CNN-TDNNF}} &
\multicolumn{1}{c}{\textbf{A}} & 
\multicolumn{1}{c}{\textbf{B}} & 
\multicolumn{1}{c}{\textbf{C}} & 
\multicolumn{1}{c}{\textbf{D}} & 
\multicolumn{1}{c}{\textbf{Avg.}} \\
\midrule
\multirow{2}{*}{-SpecAug} & Baseline & 1.55 & 3.28 & 3.03 & 9.49 & \textbf{5.80} \\
\cmidrule{2-7}
& Proposed & 1.85 & 3.57 & 2.97 & 9.32 & 5.87 \\
\cmidrule{1-7}
\multirow{2}{*}{+SpecAug} & Baseline & 1.31 & 2.27 & 1.89 & 5.74 & 3.66 \\
\cmidrule{2-7}
& Proposed & 1.23 & 2.13 & 2.02 & 5.61 & \textbf{3.55} \\
\bottomrule
\end{tabular}
\vspace{-5pt}
\end{table}

\begin{table}[t!]
\centering
\caption{State-of-the-art benchmarks on Aurora-4 with respect to bigram and trigram language models used for decoding.}
\vspace{-5pt}
\label{tab:best}
\begin{tabular}{@{\hspace{0pt}}c@{\hspace{6pt}}c@{\hspace{10pt}}c@{\hspace{10pt}}c@{\hspace{10pt}}c@{\hspace{10pt}}c@{\hspace{10pt}}c}
\toprule
&
\multicolumn{1}{c}{\textbf{Method}} &
\multicolumn{1}{@{\hspace{0pt}}c}{\textbf{A}} & 
\multicolumn{1}{@{\hspace{0pt}}c}{\textbf{B}} & 
\multicolumn{1}{@{\hspace{0pt}}c}{\textbf{C}} & 
\multicolumn{1}{@{\hspace{0pt}}c}{\textbf{D}} & 
\multicolumn{1}{@{\hspace{0pt}}c}{\textbf{Avg.}} \\
\midrule
\midrule
\multirow{2}{*}{\shortstack[1]{bi-\\gram}} & CAT-VDCRN \cite{Qian2018} & 2.95 & 4.17 & 3.70 & 8.55 & 5.92 \\
\cmidrule{2-7}
& Proposed & 2.20 & 3.42 & 3.34 & 7.36 & \textbf{5.02} \\
\midrule
\multirow{2}{*}{\shortstack[1]{tri-\\gram}} & CNN-Raw \cite{Loweimi2020} & 2.70 & 4.40 & 4.00 & 6.40 & 5.10 \\
\cmidrule{2-7}
& Proposed & 1.14 & 2.14 & 1.91 & 5.25 & \textbf{3.39} \\
\bottomrule
\end{tabular}
\vspace{-10pt}
\end{table}

First, we compared four types of aggregation functions for the predicted noise features in Fig. \ref{fig:structure} on Aurora-4. Table \ref{tab:agg} shows that STAT outperforms the other three functions, presumably because statistical information can more consistently describe the noise pattern in noisy speech. Therefore, STAT will be used in the following experiments.

Then, we evaluated the WERs of different AMs on Aurora-4. Note that our proposed model can be regarded as a combination of the baseline and SENAN. Several observations can be made from Table \ref{tab:aurora4}. First, when using TDNN-F as AM, with or without SpecAug, our proposed model outperforms the baseline in most test subsets. The average WER is reduced by 8.72\% (from 6.42\% to 5.86\% with SpecAug) and 3.23\% (from 4.03\% to 3.90\% without SpecAug). Although the WER reduction is relatively small when SpecAug is used, the system combining SpecAug and SENAN still achieves the best performance (WER=3.90\%). The results show that the two data augmentation techniques can complement each other. Second, when CNN-TDNNF is used as AM but without SpecAug, our proposed model performs slightly worse than the baseline because of its weak performance in Sets A and B. The results imply that our proposed model may underfit the training set, which was recorded using the same Sennheiser microphone as Sets A and B. However, when viewed optimistically, our proposed model is still better than the baseline in terms of handling real situations, as shown by the results of Sets C and D. Among all models, the proposed model combining CNN-TDNNF, SENAN, and SpecAug achieves the lowest WER of 3.55\%. Third, in the oracle experiment, the average WER reached 1.61\%, where the estimated clean speech and noise are replaced by ground truth. This result confirms that we are on the right track, but our model still has considerable room for improvement.

Next, we compared our best model (CNN-TDNNF+SENAN+SpecAug) with the best methods in the literature for the Aurora-4 task. Following the practice of these methods, the forced alignment of noisy speech before training neural AMs was directly obtained from that of the corresponding \textit{clean} speech. The results in Table \ref{tab:best} confirm the superiority of the proposed model. Note that the subtle WER difference (3.55\% in Table \ref{tab:aurora4} vs. 3.39\% in Table \ref{tab:best}) is due to the different forced alignment of noisy training speech.

Finally, we evaluated the proposed model on the more challenging corpus AMI. The results in Table \ref{tab:ami} again demonstrate that both SpecAug and SENAN can improve recognition performance, and they complement each other. It is also worth noting that standing on the shoulders of SpecAug, SENAN can further substantially reduce the WER on the Dev set ($4.17\%-1.19\%=\boldsymbol{2.98\%>1.19\%}=1.19\%-0\%$).

\subsection{Ablation Studies}

Our proposed model contains two key components: SENAN and the aggregation function. SENAN is reduced to a denoising autoencoder if the predicted noise feature $\mathbf{y}_{nse}$ in Fig. \ref{fig:structure} is discarded. We conducted a series of ablation studies by adding or changing network components to demonstrate their efficacy. The results in Table \ref{tab:ablation} show that each component does contribute to the performance.

\begin{table}[t!]
\centering
\caption{WERs (\%) and relative changes over Baseline on AMI.}
\vspace{-5pt}
\label{tab:ami}
\begin{tabular}{lcccc}
\toprule
\multicolumn{1}{c}{\textbf{Method}} &
\multicolumn{1}{c}{\textbf{Dev}} &
\multicolumn{1}{c}{rel. \%} &
\multicolumn{1}{c}{\textbf{Eval}} &
\multicolumn{1}{c}{rel. \%} \\
\midrule
\midrule
CNN-TDNNF (Baseline) & 33.6 & - & 37.0 & - \\
+ SpecAug & 33.2 & 1.19 & 35.9 & 2.97 \\
+ SENAN (Proposed) & \textbf{32.2} & 4.17 & \textbf{35.5} & 4.05 \\
\bottomrule
\end{tabular}
\vspace{-5pt}
\end{table}

\begin{table}[t!]
\centering
\caption{Ablation studies on Aurora-4, where AGG is the aggregation function, and ``+'' means accumulating.}
\vspace{-5pt}
\label{tab:ablation}
\begin{tabular}{@{\hspace{0pt}}lcc}
\toprule
\multicolumn{1}{c}{\textbf{Method}} &
\multicolumn{1}{c}{\textbf{WER \%}} &
\multicolumn{1}{c}{rel. \%} \\
\midrule
\midrule
TDNNF (Baseline) & 6.42 & - \\
+ SENAN (enhanced only) & 6.17 & 3.89 \\
+ SENAN (enhanced only w/ AGG) & 6.08 & 5.30 \\
+ SENAN (enhanced \& noise w/ AGG) & 5.86 & 8.72 \\
+ SpecAug & 3.90 & 39.25 \\
+ Replaced with CNN-TDNNF & \textbf{3.55} & 44.70 \\
\bottomrule
\end{tabular}
\vspace{-10pt}
\end{table}

\section{Conclusion}

In this paper, we have proposed an architecture that combines speech enhancement and acoustic modeling to achieve robust ASR and a joint training mechanism that involves ASR-related losses, in particular the lattice-free MMI, and feature reconstruction errors. The key component, SENAN, plays an important role in separately extracting enhanced speech and noise features from noisy speech. Through appropriate aggregation functions, the two features extracted by SENAN can be combined with the noisy speech feature to boost the ASR performance. They enable acoustic models to filter out non-speech content from noisy speech in a more effective way than conventional SE plus AM frameworks. Our experimental results have shown that the proposed models outperform the baseline models without SENAN on the Aurora-4 and AMI tasks, and outperform the state-of-the-art methods on the Aurora-4 task.

\bibliographystyle{IEEEtran}
\bibliography{references.bib}
\end{document}